# Deriving Application Level Relationships by Analysing the Run Time Behaviour of Simple and Complex SQL Queries

GIRISH SUNDARAM
Technical Solution Architect
IBM India Software Labs
Email: gisundar@in.ibm.com
and
MUDIT BACHHAWAT
Indian Institute of Technology
Kharagpur, India
Email: mudit.bachhawat@iitkgp.ac.in

___________________________________________________________________________

This paper describes a unique approach to perform application behavioral analysis for identifying how tables might be related to each other. The analysis techniques are based on the properties of primary and foreign keys and also the data present in their respective columns. We have also implemented the idea using JAVA and presented experimental results in Demo Section

___________________________________________________________________________

## 1. INTRODUCTION

In database management, relationships help to prevent redundant data and maintain the integrity of the database. While creating a database, we must create tables which has some physical entity. For example customer, orders, items, etc. But we also need to create relations to relate those tables. For instance, customers make orders, and orders contain items. These relationships need to be represented in the database. So for representing relationships we have the concept of primary key and foreign key.

A table is defined as a group of columns. Primary Key is a column of the table which uniquely defines each row in the table. Primary key enforces data integrity in the table. At the same time, a foreign key is a column whose value points to primary key values of another table. Foreign Key enforces the link between two tables and hence restricts the data that can be entered. It prevents user to insert invalid data.

Primary and Foreign Keys define relationships between different tables. In SQL, we use different constrained statements to carry out operations between these tables. It is not necessary to define a primary and foreign key in the database to use constrained SQL

Queries, but the main reason for primary and foreign keys is to enforce data consistency. For large databases primary and foreign keys increase the efficiency of the SQL Queries, prevents data redundancy and also prevents invalid data.

This paper introduces a unique approach to predict the possible application level relationships in databases with the help of the application relationship analysis of simple and complex SQL queries. Complex SQL queries are those which contain multiple constraints at different levels of the database. In the process of deriving relations, we first parse the SQL statements and then analyse the parsed information to extract related information.

## 2. TERMINOLOGIES
### 2.1 SQL Aliases

Aliases makes SQL operations more readable, more flexible and makes SQL simpler to understand. It also compresses the SQL code and minimizes redundant actions. Basically, SQL aliases are used to give a database table, or a column in a table, a temporary name which can be further used anywhere in the query and will be displayed in the resulting query. Mainly, Two types of aliases are used, i.e. Table Alias and Column Alias.

*2.1.1 Table Alias.* For this query:
```
SELECT *
FROM    (SELECT *
         FROM    salaries
         ORDER  BY to_date DESC) sa,
        Employees e
WHERE   e.Emp_no = sa.Emp_no
GROUP   BY sa.Emp_no;
```

In above example you can see, how table alias are used to make the code much simpler. As shown, table aliases are also used for assigning nested SQL queries.

*2.1.2 Column Alias.* In the given example, column alias is used to rename a column from functional name to a readable.
```
SELECT Customername,
       CONCAT(Address, ",", City, ",",
```

```
        Postalcode, ",", Country)
        AS Address
FROM    Customers;
```

2.4 Row Hit Rate

In Database relation, Row Hit Rate of Column A is defined to be the rate of the number of rows which are related with Column B in another table. In other words, Row hit rate represents the number of Rows in **Source Table** which has a corresponding value in **Mapped Table**. Row Hit Rate have a different value for Source and Mapped Table. Given figure will show the difference.

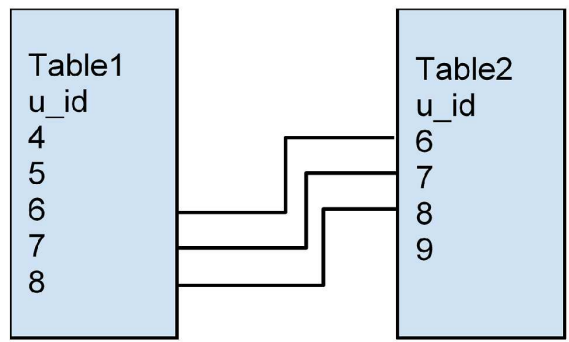

Fig. 1. Tables explaining row hit rate.

From Fig. 1, Table1 has 3 rows which are related to Table2, hence Row Hit Rate for Table1.u_id corresponding to Table2.u_id will be 3/5 = 0.6. Similarly, for Table2, Row Hit Rate will be 3/4 = 0.75

This column is represented in percentage or fraction.

2.5 Cardinality

In SQL, Cardinality represents the uniqueness of the data values contained in a table. The cardinality of a column represents the total number of unique value in the column. In Database optimization, Cardinality can be of three types, Low Cardinality, Normal Cardinality and High Cardinality.

2.6 Selectivity Rate

Selectivity rate represents the rate of unique data of column. Hence, it is defined as Cardinality divided by row count. Selectivity rate refers to the degree of uniqueness of the data. So, Higher the selectivity rate ( -> 1 ), lower will be the number of repeated rows and vice versa.

## 3. METHODS

The approach of this paper aims to predict database relationship between various tables. This whole process is divided into two modules, **SQL Parser** and **SQL Analyser**. First module i.e. SQL Parser performs **syntactic analysis** of the given SQL query and passes the required information to SQL Analyser. SQL Analyser compiles the required information to the list of the possible database relations.

### 3.1 SQL Parser

The main work of the module is to extract the relevant information from the SQL Queries. Parsing is a process of breaking a string into smaller chunks by following a set of rules so that it can be more easily interpreted, managed, or transmitted by a computer. Generally, the process of parsing is initiated by a separate **lexical analysis** which generates **tokens** from the given string. The generated tokens are further passed for **semantic analysis** to generate a **parse tree**. (This given process of the parsing is the most used process of parsing. One can create much simpler parser also according to their need). Fig. 2 explains the general process of parsing.

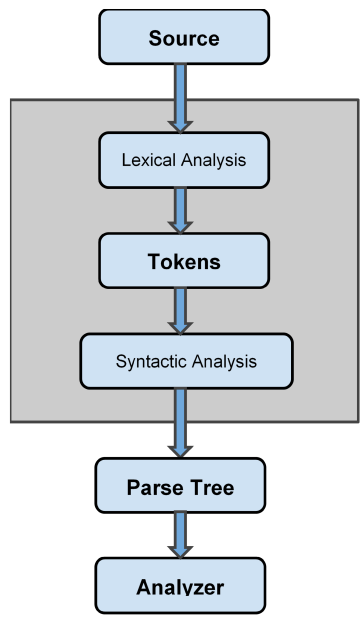

Fig. 2. Process of Parsing

We have used **ANTLR v4** as parsing tool. It's widely used to build languages, tools, and frameworks.

ANTLR basically compiles grammar rules to **target language** (such as C++, Java, etc.) and then you can include the code generated in your project. The previous version of ANTLR supported a large number of target languages. It also provides functionality to walk through the parse tree.

In this project, we need to extract two major components from an SQL query, i.e. Alias and Constraints. For example:

Consider the following SQL statement:

```
SELECT  emp.empname,
        dept.deptname
FROM    emp,
        dept
WHERE   emp.deptno = dept.deptno;
```

Fig. 3 shows the tree created by ANTLR v4. While walking through tree, the parser will extract following information:

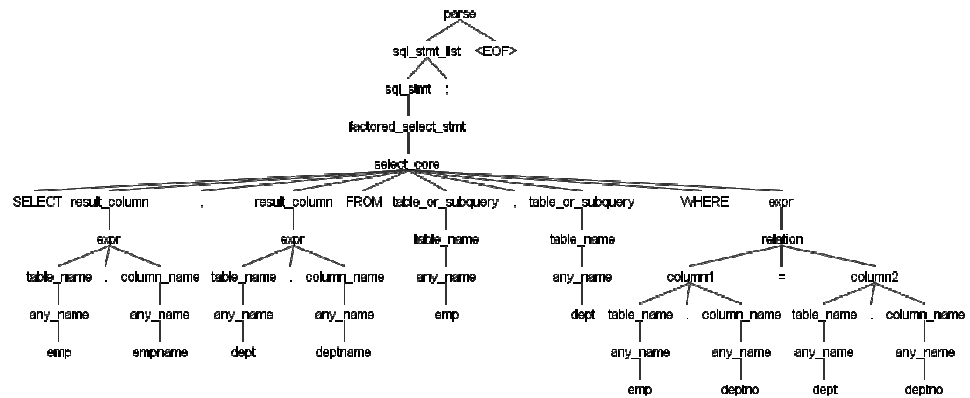

Fig. 3. Parse Tree

**Constraints**
- emp.deptno = dept.deptno

**Columns Aliases**
- None

**Table Aliases**
- None

Aliases also play a vital role in complex SQL query. In real life problems, Aliases are mainly used simplify the SQL queries.

Let's take another example:

```
SELECT *
FROM    employees e
        JOIN salaries s
          ON e.emp_no = s.emp_no
        JOIN titles t
          ON e.emp_no = t.emp_no
        JOIN dept_emp de
          ON e.emp_no = de.emp_no
        JOIN departments d
          ON d.emp_no = t.emp_no
        JOIN dept_manager dm
          ON d.dept_no = dm.dept_no
        JOIN employees edm
          ON dm.emp_no = edm.emp_no
```

[The tree of above query will be very complex, hence not displayed.]

In above example, we can easily see the importance of alias in finding a relationship. The parser will forward following details to the analyser:

**Constraints**

- e.emp_no = s.emp_no
- e.emp_no = t.emp_no
- e.emp_no = de.emp_no
- d.emp_no = t.emp_no
- d.dept_no = dm.dept_no
- dm.emp_no = edm.emp_no

**Columns Aliases**

- None

**Table Aliases**

- e - employees
- s - salaries
- t - titles
- de - dept_emp

- d - departments
- dm - dept_manager
- edm – employees

Now these parameters are passed into SQL analyser for further analysis. SQL Analyser will further validate, analyse and extract results from this information.

3.1 SQL Analysis

This module collects the parsed information and does more analysis to define relations. Initially, this module connects with database management system and validates the columns given. The process of validation includes verifying the existence of columns, use of aliases (if any), extracting the current information of data type of columns, etc. and output the final list of columns and tables which might be related to each other.

Once validation is completed it evaluates Row Hit Rate and Selectivity Rate for each of the respective relations for further classification into primary and foreign key.

Table I shows properties of a primary and foreign key on ideal condition

| Primary Key | Foreign Key |
|---|---|
| Primary key is used to uniquely identify a record in the table. | A Foreign key is a field in the table that is a primary key in another table. i.e. Foreign key values can be repeated. |
| Primary Key can't accept null values. | A Foreign key can accept multiple null values. |

Table I. Properties of Primary and Foreign Key

Since we know that primary key is always unique but foreign key may be repeated hence with the help of Selectivity Rate we can tell that primary key will have a higher selectivity rate.

Let's say, we have a constraint with Selectivity rate as follows:

Column 1 Selectivity Rate - 99% (1486/1500)
Column 2 Selectivity Rate - 3% (1456/48000)

Then it is quite obvious that Column 1 should be the primary key and primary column 2 is the foreign key. But many exceptional cases may also arise in which we cannot apply this formula. Such as:

Column 1 Selectivity Rate - 99% (1486/1500)

Column 2 Selectivity Rate - 100% (99/99)

Here since the number of values in column 2 is less than Column, then It may happen that higher selectivity rate in Column 2 is because of few number of entries related to Column 1. In such cases, the decision can be made on the basis of Row Hit Rate. Row hit rate is the rate of number rows which are related with the constraints. Hence, Foreign Key will have a higher Row Hit Rate. Hence with this concept, Column 1 will be the primary key. Hence, we can set a threshold for the number of columns and for the percentage to decide which rule will be applicable.

## 4. CALCULATION

In this section we have explained the evaluation of various components which were used in SQL Analyser.

### 4.1 Finding Selectivity Rate

Selectivity rate is defined as

$$Selectivity\ Rate = \frac{Number\ of\ unique\ values\ in\ Column}{Total\ Number\ of\ values\ (rows)}$$

Hence, for a given column, we can find total number of rows as

```
SELECT Count(*) x
FROM    <tablename>
```

And similarly total number of unique rows with

```
SELECT Sum(u) x
FROM    (
            SELECT Count(v) u
            FROM    (
                        SELECT   (Count(*)) v
                        FROM     <tablename>
                        GROUP BY <columnname>) p) q;
```

### 4.1 Finding Row Hit Rate

$$Row\ Hit\ Rate = \frac{Number\ of\ rows\ related\ to\ another\ table}{Total\ Number\ of\ values\ (rows)}$$

The total number of rows related to another table can be calculated as follows,

```
SELECT    Count(*) count
FROM      <table1>
LEFT JOIN <table2>
ON        <table1>.<column1> = <table2>.<column2>
WHERE     <table1>.<column1> IS NOT NULL
AND       <table2>.<column2> IS NULL
```

Here table1 and column1 are the details of table and column whose Row Hit Rate is to be calculated and table2 and column2 are the details of the table and column which might be related to table1 respectively.

## 5. DEMONSTRATION

For demonstration process, we will be using employee database from [here](). Fig 4 shows the structure of the database.

Fig. 4. Database Structure

For experiment purpose, we have taken a large number of rows, to measure a significant difference. This table contains:

| Table Name | Number of Rows |
|---|---|
| departments | 9 |
| employees | 300024 |
| dept_emp | 331603 |
| dept_manager | 24 |
| salaries | 2844047 |
| titles | 443308 |

So, let's process the queries give below with our application.

```
SELECT *
FROM    employees
    INNER JOIN (SELECT *
            FROM    (SELECT *
                    FROM    salaries
                    ORDER   BY emp_no,
                            from_date DESC) temp
            GROUP   BY emp_no) tp
        ON employees.emp_no = tp.emp_no
    INNER JOIN (SELECT *
            FROM    (SELECT *
                    FROM    dept_emp
                    ORDER   BY emp_no,
                            from_date DESC) temp2
            GROUP   BY emp_no) tt
        ON tt.emp_no = employees.emp_no;
```

The program will individually process these entries. SQL Parser will parse these queries and return following parameters for SQL Analyser.

**Analysis**

**Constraints**

- employees.emp_no = tp.emp_no
- tt.emp_no = employees.emp_no

**Column Alias**

- None

**Table / SQL Alias**
- temp - SELECT * FROM salaries ORDER BY emp_no, from_date DESC
- tp - SELECT * FROM temp GROUP BY emp_no
- temp2 - SELECT * FROM dept_emp ORDER BY emp_no,from_date DESC
- tt - SELECT * FROM temp2 GROUP BY emp_no

Now once these parameters are extracted, analyser will backtrace the actual column from the SQL Alias and will then validate the table information from the database. And then give the actual proper constraints.

**Constraints**
- employees.emp_no = salaries.emp_no
- dept_emp.emp_no = employees.emp_no

Based on this constraints we can infer that there might be a potential relationship between employee - salaries and dept_emp - employees tables on the emp_no column. So using this concept we can now extrapolate this requirement all kind of SQL statements. Once this information is finalized then analyser will further extract its column information and then appropriate PK and FK relation.

| Column | Row Hit Rate | Selectivity Rate |
|---|---|---|
| employees.emp_no | 100% (300024/300024) | 100% (300024/300024) |
| salaries.emp_no | 100% (2844047/2844047) | 11% (300024/2844047) |

In above example, we can see that selectivity rate for employee is higher that salaries i.e. salaries.emp_no will have repeated entries from employee.emp_no. Hence, employee.emp_no will be primary key and salaries.emp_no will be foreign.

For second constraint,

| Column | Row Hit Rate | Selectivity Rate |
|---|---|---|
| dept_emp.emp_no | 100% (331603/331603) | 90% (300024/331603) |
| employees.emp_no | 100% (300024/300024) | 100% (300024/300024) |

Similarly, In this case also, we can see that employee.emp_no will be primary key and salaries.emp_no will be foreign. Hence, the program will generate a csv file containing all the details of the relations.

6. DRAWBACK OF EXISTING SOLUTIONS

Till now, we have seen that how this program generates a list of possible relations with the help of few queries. IBM Infosphere Discovery also performs database relations prediction for simple to complex systems. But this solution provides the same operations with a high reduction in space and time, which is great when the database is complex and large.

The process in IBM Infosphere Discovery which manages the data and examine relationships is named as the discovery process. Discovery process automatically analyses the data and discover the relationships with the help of input from the data analyst. The main work of the analyst is to validate the discovered results and to select the most appropriate options to be implemented in the database. This discovery part is divided into 2 parts:

**Analysis Process**: The analysis process is based on the identification of data types and then discovering the relations with each possible dataset (Source and Target). Basically, it just looks for the data values matches between source and target tables and create a list of all such possible combination. IBM Infosphere automatically performs these actions, but finally for best results the results are filtered by a data analyst.

In the Fig. 5 for the given table structure, RED marked relations are "false" relations and need to be filtered.

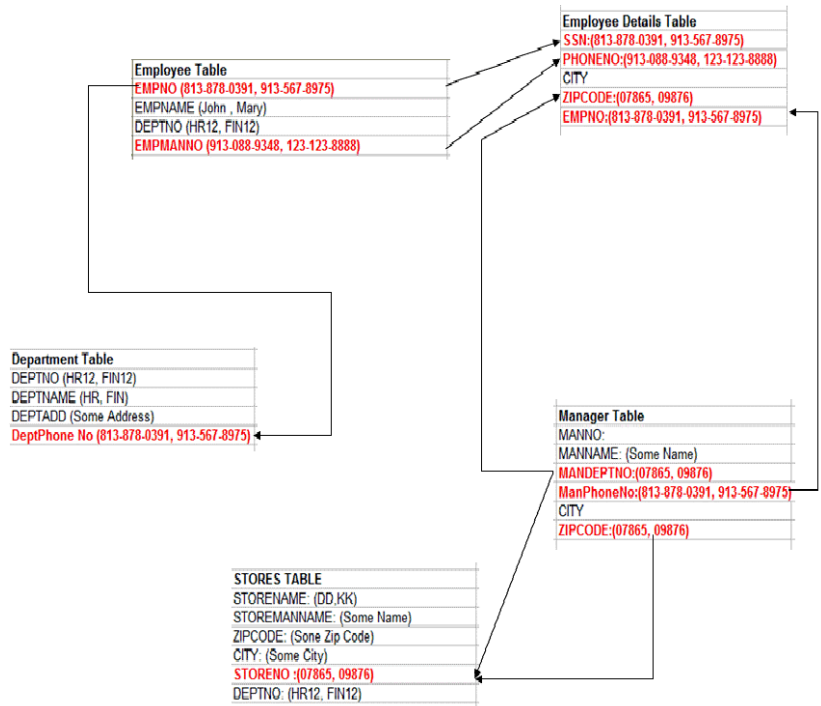

Fig. 5. False Relations

**Mapping Process**: The mapping process uses the discovered results to further discover the joins, bindings, and transformations that correctly derive the target data from source data. Again, Infosphere Discovery performs this action automatically, but to achieve the highest possible accuracy, you need to review and analyse the result.

**Summary of the drawback**
- **Require Human Intervention**: At different stages of discovery, the results need to be verified and filtered by an analyst to get the best possible solutions and highest possible accuracy.
- **Analysis of the False Relationships**: If the relationships are not filtered by the analyst then there might be situations where a large number of false queries will also be evaluated for mapping process resulting in unnecessary time consumption.
- **Time and Space Complexity**: It can be said that the analysis performed by discovery process is similar to cross join, which creates the list of all possible values. And hence for large datasets, discovery process not only takes more time but also consume a lot of resources. For example, if we have a single query on two tables with let's say 5 columns each. If for the proposed idea requires 1 unit of time this existing technique will be taking 25 units of time and number of rows times increment in space complexity.

- **The Need of Physical Understanding of Database Structure**: For data analyst to filter the relationships, they need a physical interpretation of the database and then convert those interpretations to relations.

From the above points, it is very clear that actual physical data analysis to derive database relationships is very inefficient in all aspects. So, there is a need of an intelligent autonomic methodology to identify application level relationships by analysing the run-time behaviour so that we can accurately and efficiently get the insights of how various business entities / tables / data models are related to each other.

7. RESULT

Till now we have seen that how can we predict database relations with the help of properties and data present in the database. Of Course a static analysis will not be able to gather all the information, so a dynamic approach is needed. The dynamic approach can be used in addition with static analysis to optimize all the information. During runtime, the above program will act as a middleware component between the database system and SQL runner or else we can create a regular monitor to monitor all the query in a certain interval of time. This will again help to gather more information regarding relationships. But even with the static and dynamic approach, all relationships will not be detected, but discovery process will be substantially improved.

On the other side, discovering relationships will drastically improve the performance. Assigning relationships the database will help to maintain data integrity. Defining relationship can also gain up to 90% reduction in time with ease of debugging in various data problems. For big data systems, optimizing the database is an important issue which must be tackled. Below are the few comparisons of the queries performed with optimization

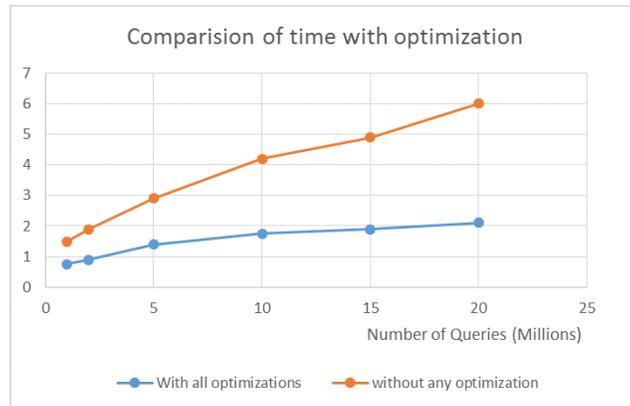

Fig. 6. Showing time vs number of query for with optimization and without optimization